\documentclass[a4paper]{JHEP}
\usepackage[T1]{fontenc}


\newcommand{\LyX}{L\kern-.1667em\lower.25em\hbox{Y}\kern-.125emX\spacefactor1000}
\title 
{A \( \kappa  \)\( \;  \)Gauge Fixed Type IIB Superstring Action on \( AdS_{5}\times S_{5} \)} 

\author
{
 I. Pesando \\
 Dipartimento di Fisica Teorica , Universit\'a di Torino, via P. Giuria 1,
 I-10125 Torino, Istituto Nazionale di Fisica Nucleare (INFN) - sezione di 
 Torino, Italy
\thanks
{
 email:ipesando@to.infn.it, pesando@alf.nbi.dk
}
\thanks
{ 
 Work supported by the European Commission TMR programme
 ERBFMRX-CT96-004
}
}

\abstract{ We describe the type IIB supergravity background on \(
  AdS_{5}\times S_{5} \)using the potentials of \( AdS_{5|4}\bigotimes
  S_{5} \) and we use the supersolvable algebra associated to \(
  AdS_{5|4} \) to compute the \( \kappa \) gauge fixed type IIB string
  action.  }

\preprint
{ DFTT-45/98 \\ hep-th/9808020}

\csname @addtoreset\endcsname{equation}{section}

\begin{document}

\section{Introduction.}

It has become clear in the last period that in order to exploit the celebrated
conjecture on the AdS/SYM\cite{Maldacena}\cite{GubserKlenanovPolyako}\cite{Witten}
correspondence fully is of the utmost importance to be able to compute string
amplitudes on \( AdS_{5}\times S_{5} \) background\cite{BanksGreen}\cite{Li}:
this would be equivalent to control \( N=4 \) SYM in the large \( N \) limit
for all the possible values of the 't Hooft coupling constant \( \lambda =g^{2}_{YM}N \).

The problem has been tackled constructing the \( \sigma  \)-model associated
with the supergroup \( \frac{SU(2,2|4)}{SO(1,4)\bigotimes O(5)} \) \cite{MatsaevTseytlin}\cite{Kallosh}
which is the global symmetry of the background; in order to be able to use the
resulting action one has first of all to fix the \( \kappa  \)-symmetry and
until very recently \cite{Kall-fixing} there was no proposal on how to do it.

It is purpose of this letter to obtain such a \( \kappa  \) gauge fixed action
but in order to achieve this we will use a little different approach and we
will follow the supersolvable approach developed in \cite{gruppoTo}.

This approach consists in the following steps:

\begin{enumerate}
\item find the field strengths and the reonomic parameterization corresponding to the
\( AdS_{5}\times S_{5} \) background;
\item write the IIB superspace background in terms of the potentials of \( AdS_{5|4}\bigotimes S_{5}\equiv \frac{SU(2,2|4)}{SO(1,4)\bigotimes U(4)}\bigotimes \frac{SO(6)}{SO(5)} \)
( which allows moreover to easily generalize to more general backgrounds);
\item find the supersolvable algebra associated with \( AdS_{5|4} \). In doing so
we project out \( \frac{1}{2} \) of the fermionic coordinates thus fixing a
gauge for the \( \kappa  \) symmetry; 
\item insert the previous parameterization in the appropriate action which , in this
case, is the one computed by Grisaru, Howe, Mezincescu, Nilsson and Townsend
\cite{GHMNT}.
\end{enumerate}
We will now sketch these steps and give the results of each of them leaving
for a future publication a more complete explanation along with the application
to the \( D3 \) case \cite{D3-to} .

\section{Step 1: determining the superbackground.}

Using the notation of \cite{IIB}\footnote{
With the minor modifications given by \( \psi \rightarrow \Psi  \) , \( \psi ^{*}\rightarrow \Psi _{c}\equiv \widehat{C}\overline{\Psi }^{T} \)and
\( \varpi ^{ab}\rightarrow \Omega ^{\widehat{a}\widehat{b}} \).
}it is an easy matter to find the field strengths associated with the \( AdS_{5}\times S_{5} \)
solution:

\begin{eqnarray}
R^{ab}_{..\, cd}=-e^{2}\, \delta ^{ab}_{cd} & \;  & R^{ij}_{..\; kl}=e^{2}\, \delta ^{ij}_{kl}\label{eq-mot-Riem} \\
F_{abcde}=\frac{e}{5}\epsilon _{abcde} & \,  & F_{ijklm}=\frac{e}{5}\epsilon _{ijklm}\\
 & F^{\alpha }_{\widehat{a}\widehat{b}\widehat{c}}=0 & \\
 & P_{\widehat{a}}=0 & \\
 & D_{\widehat{a}}\lambda =0 & \\
 & \rho _{\widehat{a}\widehat{b}}=0 & \label{eq-mot-grav} 
\end{eqnarray}
where \( \widehat{a},\widehat{b},\ldots =0\ldots 9 \) , \( a,b,\ldots =0\ldots 4 \)
and \( i,j,\ldots =5\ldots 9 \).

\section{Step 2: expressing the superbackground through the \protect\( AdS_{5|4}\bigotimes S_{5}\protect \)
generalized potentials.}

We want now to express the IIB generalized potentials \( \left\{ \omega ^{\widehat{a}\widehat{b}},V^{\widehat{a}},\Psi ,V^{\pm }_{\alpha }A^{\alpha },B\right\}  \),
thought as fields living on the \( AdS_{5|4}\bigotimes S_{5} \) superspace,
using a linear combination of the following fields:

\begin{itemize}
\item the generalized potentials describing the 5D super anti de Sitter \( AdS_{5|4}=\frac{SU(2,2|4)}{SO(1,4)\bigotimes U(4)} \)
supergeometry  \( \left\{ \omega ^{ab}(x,\Theta )\: ,\: P^{a}(x,\Theta )\: ,\: A^{N}_{M}(x,\Theta )\: ,\: B(x,\Theta )\: ,\: \psi _{N}(x,\Theta )\right\}  \)
(\( N=1,\ldots ,4 \));
\item the funfbein and spin connection \( \left\{ E^{i}(z),\omega ^{ij}(z)\right\}  \)
describing the classical geometry of \( S_{5}=\frac{SO(6)}{SO(5)} \).
\end{itemize}
In order to proceed and enable the reader to reproduce the computations we need
define in a more precise manner the previous mentioned fields:

\begin{itemize}
\item the generalized potentials of the 5D super anti de Sitter superspace \\
\( AdS_{5|4} \) \( \left\{ \omega ^{ab}(x,\Theta )\: ,\: P^{a}(x,\Theta )\: ,\: A^{N}_{M}(x,\Theta )\: ,\: B(x,\Theta )\: ,\: \psi _{N}(x,\Theta )\right\}  \)
satisfy the \( su(2,2|4) \) Maurer-Cartan equations :
\begin{eqnarray}
d\varpi ^{ab}-\varpi ^{ac}\varpi _{c}^{.b}+e^{2}P^{a}P^{b}+\frac{e}{2}\overline{\psi }^{N}\gamma ^{ab}\psi _{N} & = & 0\label{MC-omega} \\
dP^{a}-\varpi ^{a}_{.b}P^{b}-\frac{i}{2}\overline{\psi }^{N}\gamma ^{a}\psi _{N} & = & 0\\
dB+\frac{e}{2}\overline{\psi }^{N}\psi _{N} & = & 0\\
dA^{N}_{.M}+eA^{N}_{.P}A^{P}_{.M}+\overline{\psi }^{N}\psi _{M} & = & 0\\
d\psi _{M}-\frac{1}{4}\varpi ^{ab}\gamma _{ab}\psi _{M}+\frac{ie}{2}P^{a}\gamma _{a}\psi _{M} & +e\psi _{N}A^{N}_{.M} & =0\label{MC-psi} 
\end{eqnarray}
where \( \overline{\psi }=\psi ^{\dagger }\gamma _{0},\, \psi _{c}=C\overline{\psi }^{T} \)
and depend on \( \left\{ x^{a},\Theta _{N}\right\}  \) \footnotemark{}
 (\( N=1,\ldots ,4 \)) where the fermionic coordinates \( \Theta _{N}=\left( \begin{array}{c}
\theta _{N}\\
\xi _{N}
\end{array}\right)  \) transform as spinors \emph{only} with respect to \( SO(1,4) \) and \emph{not}
with respect to \( SO(1,4)\otimes SO(5) \) : they are a set of 4 (symplectic)
complex spinors in 5D which are the result of the dimensional reduction of two
Majorana-Weyl 10D spinors. The Maurer Cartan equations (\ref{MC-omega}-\ref{MC-psi})
are derived from the generic 1-form valued Lie algebra element
\begin{equation}
\label{mu-su22N}
\mu =\left( \begin{array}{cc}
-\omega ^{ab}\frac{1}{4}\gamma _{ab}+eP^{a}\frac{i}{2}\gamma _{a} & \sqrt{e}\, \psi _{M}\\
\sqrt{e}\, \overline{\psi }^{N} & A^{N}_{.M}
\end{array}\right) +B\: 1_{8}
\end{equation}
 which is defined by the condition 
\begin{equation}
\label{def-su22N}
\mu ^{\dagger }G+G\mu =0\; \; \; \; G=\left( \begin{array}{cc}
-\gamma _{0} & \\
 & 1_{N}
\end{array}\right) 
\end{equation}
This condition allows also to write the generic element of the \( su(2,2|4) \)
superalgebra as 
\begin{equation}
\label{g-su22N}
g=\left( \begin{array}{cc}
t^{ab}\frac{1}{4}\gamma _{ab}+t^{a}\frac{i}{2}\gamma _{a} & \Theta _{M}\\
\overline{\Theta }^{N} & t^{N}_{M}
\end{array}\right) 
\end{equation}
and write the generic element of vector space associated with the coset \( AdS_{5|4} \)
either in the usual way as
\begin{equation}
\label{g-su22N-usual}
l=\left( \begin{array}{cc}
t^{a}\frac{i}{2}\gamma _{a} & \Theta _{M}\\
\overline{\Theta }^{N} & 0
\end{array}\right) 
\end{equation}
or using the solvable algebra associated with \( AdS_{5} \) as 
\begin{equation}
\label{g-su22N-solv}
l=\left( \begin{array}{cc}
t^{4}\frac{i}{2}\gamma _{4}+t^{p}\frac{i}{2}\gamma _{p}\frac{1+i\gamma _{4}}{2} & \sqrt{e}\, \Theta _{M}\\
\sqrt{e}\, \overline{\Theta }^{N} & 0
\end{array}\right) 
\end{equation}
 and compute a coset representative of the superspace exponentiating the previous
algebra element.
\item the classical funfbein and spin connection \( \left\{ E^{i}(z),\omega ^{ij}(z)\right\}  \)
of \( S_{5}=\frac{SO(6)}{SO(5)} \) endowed with the killing induced metric,
i.e. negative definite, satisfy 
\begin{eqnarray}
dE^{i}-\varpi ^{i}_{.j}E^{j} & = & 0\label{S5-tors} \\
d\varpi ^{ij}-\varpi ^{ik}\varpi _{k}^{.j} & = & e^{2}E^{i}E^{j}\label{S5-Riem} 
\end{eqnarray}
 and depend on the coordinate \( z \). Explicitly we have ( \( z^{2}=\eta _{ij}z^{i}z^{j}=-\delta _{ij}z^{i}z^{j} \)
)
\end{itemize}
\footnotetext{
The general superalgebra \( su(2,2|M) \) is defined by the Maurer-Cartan system
(\ref{MC-omega}-\ref{MC-psi}) with the index \( N \) assuming values in \( \left\{ 1,\ldots ,M\right\}  \)
where all potentials depend on the coordinates \( \left\{ x^{ab},x^{\bullet },x^{a},\Theta _{N}\right\}  \)
(\( a,b,\ldots =0,\ldots ,4 \)) of the associated supergroup. The \( AdS_{5|M} \)
potentials satisfy the same equations but depend only on the coordinates of
the associated coset, i.e. \( \left\{ x^{a},\Theta _{N}\right\}  \).
}%

\begin{eqnarray}
E^{i} & = & \frac{2}{e}\frac{dz^{i}}{1-z^{2}}\label{S5-E} \\
\varpi ^{ij} & = & \frac{4z^{[i}dz^{j]}}{1-z^{2}}\label{S5-omega} 
\end{eqnarray}
After having properly defined the fields we are going to use we can proceed
in a similar way as done in \cite{gruppoTo} for the 11D supergravity and express
the IIB generalized potentials through the superspace potentials; we obtain

\begin{eqnarray}
V^{a}=P^{a}(x,\Theta ) & \;  & V^{i}=E^{i}(z)+\frac{1}{2}\overline{\eta }_{N}\tau ^{i}\eta ^{M}\, A^{N}_{M}(x,\Theta )\nonumber \\
 &  & \label{param-V} \\
\Omega ^{ab}=\varpi ^{ab}(x,\Theta ) & \;  & \Omega ^{ij}=\varpi ^{ij}(z)-\frac{e}{2}\overline{\eta }_{N}\tau ^{ij}\eta ^{M}\, A^{N}_{M}(x,\Theta )\nonumber \\
 &  & \\
\Omega ^{ai} & = & 0\; \; \\
\Psi  & = & \frac{1}{\sqrt{2}}\left( \begin{array}{c}
0_{16x1}\\
\psi _{N}(x,\Theta )\bigotimes \eta ^{N}(z)
\end{array}\right) \; \; \\
\epsilon _{\alpha \beta }V_{+}^{\alpha }A^{\beta } & = & -\frac{i}{e}\overline{\psi }_{cN}\psi _{M}\, \overline{\eta }_{c}^{N}\eta ^{M}\; \; \\
\left( \begin{array}{cc}
V^{1}_{-} & V^{2}_{-}\\
V^{1}_{-} & V^{2}_{+}
\end{array}\right)  & = & const\in SU(1,1)\label{param-dilat} 
\end{eqnarray}
 where we have introduced the 4 \( S_{5} \)c-number Killing spinors (\( N=1\ldots 4 \))
defined by

\begin{equation}
\label{kill-spin}
D_{SO(5)}\eta ^{N}\equiv \left( d-\frac{1}{4}\varpi ^{ij}\tau _{ij}\right) \eta ^{N}=+\frac{e}{2}\tau _{i}\eta ^{N}E^{i}
\end{equation}
 and their conjugates defined as \( \eta _{cN}=C_{5}\overline{\eta }_{N} \)
with \( \overline{\eta }\equiv \eta ^{\dagger } \) with the normalization \( \overline{\eta }_{N}\eta ^{M}=\delta ^{M}_{N} \)
.

The proof is based on application of the standard Fierz identities.

\section{Step 3: fixing the \protect\( \kappa \protect \) symmetry using the supersolvable
algebra.}

We are now ready to perform the third step: to compute the solvable superalgebra
associated with the anti de Sitter superspace \( AdS_{5|4} \), thus fixing
the \( \kappa  \) symmetry (see app. \ref{Why-kappa-fix} for a complete discussion
of the reasons why this works properly). The resulting solvable superalgebra
is described by the Maurer-Cartan equations:

\begin{eqnarray}
dP & = & 0\label{Ssolv1} \\
dN^{p}+eN^{p}P-i\frac{e}{\sqrt{2}}\overline{\chi }^{N}\gamma ^{p}\chi _{N} & = & 0\\
d\chi _{N}-\frac{e}{2}P\chi _{N} & = & 0\label{Ssolv3} 
\end{eqnarray}
where \( p,q,\ldots \in \{0,1,2,3\} \), \( P\equiv P^{4} \) , \( N^{p}\equiv \frac{\varpi ^{p4}+e\: P^{p}}{\sqrt{2}} \)
and \( \chi _{N}=\frac{1+i\gamma ^{4}}{2}\psi _{N} \) with all the other linearly
independent \( AdS_{5|4} \) potentials set to zero. From the knowledge of which
1-forms survive in (\ref{mu-su22N}) and from the explicit expression for the
generic element of the \( su(2,2|4) \) algebra (\ref{g-su22N}) we can write
the generic element of the supersolvable algebra as

\begin{equation}
\label{Ssolv-element}
g(u,y^{p},\theta )=\left( \begin{array}{ccc}
-\frac{e}{2}u\, 1_{2} & \frac{i}{\sqrt{2}}y^{p}\, \sigma _{p} & \sqrt{e\, }\theta _{N}\\
. & \frac{e}{2}u\, 1_{2} & .\\
. & \sqrt{e}\, \vartheta ^{\dagger N} & .
\end{array}\right) =u{\cal C}+y^{p}{\cal N}_{p}+\theta ^{\dagger N}Q_{N}+\overline{Q}^{N}\theta _{N}
\end{equation}

Then we can obtain an explicit expression for the potentials by using the usual
expression \( \mu =G^{-1}dG \) with \( G=\exp (u{\cal C})\, \exp (y^{p}{\cal N}_{p})\, \exp (\theta ^{\dagger N}Q_{N}+\overline{Q}^{N}\theta _{N}) \)

\begin{eqnarray}
P & = & du\label{raw-solv-P} \\
N^{p} & = & dy^{p}-ey^{p}du+i\frac{e}{2\sqrt{2}}\left( \vartheta ^{\dagger N}\widetilde{\sigma }^{p}d\theta _{N}-d\vartheta ^{\dagger N}\widetilde{\sigma }^{p}\theta _{N}\right) \\
\chi _{N} & = & \left( \begin{array}{c}
d\theta _{N}-\frac{e}{2}\theta _{N}du\\
0
\end{array}\right) \\
\overline{\chi }^{N} & = & \left( \begin{array}{cc}
0 & d\vartheta ^{\dagger N}-\frac{e}{2}\vartheta ^{\dagger N}du
\end{array}\right) \label{raw-solv-chibar} 
\end{eqnarray}
 From these expression (\ref{raw-solv-P}-\ref{raw-solv-chibar}) we extract
the \( \kappa  \) gauge fixed expression of the \( AdS_{5|4} \) potentials
as

\begin{eqnarray}
P^{p} & = & \frac{1}{e}\varpi ^{p4}=\frac{1}{e}\left( \rho dx^{p}+\frac{ie}{4}\rho \left( \vartheta ^{\dagger N}\widetilde{\sigma }^{p}d\theta _{N}-d\vartheta ^{\dagger N}\widetilde{\sigma }^{p}\theta _{N}\right) \right) \label{solv-P+Omega} \\
P^{4} & = & \frac{1}{e}\frac{d\rho }{\rho }\\
\psi _{N} & = & \sqrt{\rho }\left( \begin{array}{c}
d\theta _{N}\\
0
\end{array}\right) \\
A^{N}_{.M} & = & B=0\label{solv-A} 
\end{eqnarray}
 where we have changed variables as \( u\rightarrow \rho =e^{eu} \), \( y^{p}\rightarrow x^{p}=\frac{1}{\sqrt{2}}e^{-eu}y^{p} \)
, \( \theta \rightarrow \theta '=e^{-eu/2}\theta  \) and dropped the prime
for the new \( \theta  \) variable. 

One could wonder why this procedure fixes a gauge for the \( \kappa  \) symmetry.
The reason is very simple: it is equivalent to give a projector for the fermionic
coordinates. In fact to fix the \( \kappa  \) symmetry is roughly equivalent
to throw away half of the fermionic coordinates and then to set these coordinates
to zero in the general expression for the \( AdS_{5|4} \) potentials. All these
steps are accomplished in one step by computing the potentials for the supersolvable
algebra but we could as well have performed all these steps explicitly by first
computing the \( AdS_{5|4} \) potentials and hence the background fields as
a function of \( \left\{ x,z,\Theta _{N}=\left( \begin{array}{c}
\theta _{N}\\
\xi _{N}
\end{array}\right) \right\}  \) and then setting \( \xi _{N}=0 \) which are the fermionic coordinates projected
out by the supersolvable gauge. What one has to check is whether the projector
on the \( \Theta _{N} \) coordinates is compatible with the \( \kappa  \)
symmetry variation and with the equations of motion: this is done in app. \ref{Why-kappa-fix}
where we show that with this gauge fixing we are able to describe all string
excitations but the ones which do not satisfy
\begin{equation}
\label{kappa-descrive}
\Pi ^{p}_{\tau }\Pi _{p\tau }+\Pi ^{4}_{\sigma }\Pi _{4\sigma }+\Pi ^{i}_{\sigma }\Pi _{i\sigma }\neq 0
\end{equation}
where \( V^{\widehat{a}}=\Pi ^{\widehat{a}}_{\tau }\: d\tau +\Pi ^{\widehat{a}}_{\sigma }\: d\sigma  \).
This kind of impossibility of describing all the string excitations is similar
to what happens in the usual GS superstring in the light cone gauge on flat
superspace where we can describe all excitations but the massless ones propagating
in the + direction, the difference is that here we cannot easily understand
which states do not satisfy eq. (\ref{kappa-descrive}); we can nevertheless
find a class of such states which is made by massless states with \( \Theta =0 \)
propagating at \( r=const \) since in this case we have trivially \( \Pi ^{4}_{\sigma }=\Pi ^{i}_{\sigma }=0 \)
and \( \Pi ^{p}_{\tau }\Pi _{p\tau }=0 \).

\section{The action.}

We have now all the necessary expressions which can be inserted into the action
given app. \ref{app-action-first-order} (with \( \phi =\frac{\pi }{4} \) )
or, that is the same, in the general action given in (\cite{GHMNT}) (after
a careful comparison with \cite{IIB-HW}\footnote{
The dictionary between the notations used by Howe and West \cite{IIB-HW} and
ours \cite{IIB} is obtained setting \( B=1 \) and \( \widehat{C}=\left( \begin{array}{cc}
 & 1\\
-1 & 
\end{array}\right)  \) and it is given by:

\begin{eqnarray*}
\gamma _{0}=\Gamma _{0} & \;  & \gamma _{1,\ldots ,9}=-\Gamma _{1,\ldots ,9}\\
 & \Omega _{a}^{.b}=\Omega ^{\widehat{b}}_{\: .\widehat{a}} & \; \\
E^{a}=V^{\widehat{a}} & \;  & \left( \begin{array}{c}
E^{\alpha }\\
0
\end{array}\right) =\Psi \, e^{i\pi /4}\\
P_{HW}=P & Q_{HW}=-\frac{1}{2}Q & \left( \begin{array}{c}
0\\
\Lambda _{\alpha }
\end{array}\right) =\frac{1}{2}\lambda \, e^{-i\pi /4}\\
\;  & F_{HW}=\frac{i}{4}\epsilon _{\alpha \beta }V^{\alpha }_{+}dA^{\beta } & \; 
\end{eqnarray*}
 Notice that for this comparison we have used \( \Gamma _{11}=\left( \begin{array}{cc}
-1 & \\
 & 1
\end{array}\right)  \) instead of the more usual expression \( \Gamma _{11}=\left( \begin{array}{cc}
1 & \\
 & -1
\end{array}\right)  \) which is the one used in the paper.
} too) and get\footnote{
In the main formula (19) of \cite{GHMNT} the factor \( \frac{1}{2} \) should
be moved from the second addend to the first. This can be seen either by comparing
with the expression for the flat spacetime (1) or by looking at the expression
for the variation of the action (23).
}

\begin{eqnarray}
 & S=\int d^{2}\xi \, \sqrt{-g}\, g^{\alpha \beta }\frac{1}{2}\times  & \nonumber \\
 & \left\{ \eta _{pq}\frac{\rho ^{2}}{e^{2}}\left[ \partial _{\alpha }x^{p}+\frac{ie}{4}\left( \vartheta ^{\dagger N}\widetilde{\sigma }^{p}\partial _{\alpha }\theta _{N}-\partial _{\alpha }\vartheta ^{\dagger N}\widetilde{\sigma }^{p}\theta _{N}\right) \right] \left[ \partial _{\beta }x^{q}+\frac{ie}{4}\left( \vartheta ^{\dagger M}\widetilde{\sigma }^{q}\partial _{\beta }\theta _{M}-\partial _{\beta }\vartheta ^{\dagger M}\widetilde{\sigma }^{q}\theta _{M}\right) \right] \right. \,  & \nonumber \\
 & \left. -\frac{1}{e^{2}}\frac{\partial _{\alpha }\rho \partial _{\beta }\rho }{\rho ^{2}}-\frac{4}{e^{2}}\delta _{ij}\frac{\partial _{\alpha }z^{i}\partial _{\beta }z^{j}}{(1-z^{2})^{2}}\right\}  & \nonumber \\
 & -\frac{i}{4e}\rho \: \left[ d\theta ^{\dagger N}\sigma _{2}d\theta ^{*M}\, \overline{\eta }_{N}(z)\eta _{cM}(z)+d\theta ^{T}_{N}\sigma _{2}d\theta _{M}\, \overline{\eta }_{c}^{N}(z)\eta ^{M}(z)\right]  & \label{final-action} 
\end{eqnarray}
 This action is highly non linear and in order to be able to compute the string
scattering amplitudes we have now to solve the hardest part of the problem,
i.e. to find a way to quantize this expression, perhaps rewriting the previous
action in term of free fields.

\textbf{Acknowledgements}

It is a pleasure to thank L. Gualtieri and A. Lerda for discussions and the
NBI for hospitality during the completion of this work.

\medskip{}
\textbf{Note Added.}

\medskip{}
To request of the referee I have added appendices B and C where it is explained
why and in which way the \( \kappa  \) symmetry is fixed by this approach.

\appendix

\section{Conventions.}

\begin{itemize}
\item Indices: \( \widehat{a},\widehat{b},\ldots \in \{0,\ldots ,9\} \); \( a,b,\ldots \in \{0,1,2,3,4\} \),
\( i,j,\ldots \in \{5,\ldots ,9\} \); \( p,q,\ldots \in \{0,1,2,3\} \), \( t,u,v,\ldots \in \{4,\ldots ,9\} \);
\( \alpha ,\beta ,\ldots \in \left\{ 0,1\right\}  \) 
\item Metrics: \( \eta _{\widehat{a}\widehat{b}}=diag(+1,-1,\ldots -1) \), \( \eta _{ab}=diag(+1,-1,-1,-1,-1) \)
, \( \eta _{ij}=diag(-1,-1,-1,-1,-1) \); \( \eta _{\alpha \beta }=diag(+1,-1) \)
\item Epsilons: \( \epsilon _{0\ldots 9}=\epsilon _{0\ldots 4}=\epsilon _{5\ldots 9}=1 \);
\( \epsilon _{01}=1 \) 
\item 1+9D gamma matrices
\begin{eqnarray*}
\{\Gamma _{\widehat{a}},\Gamma _{\widehat{b}}\} & = & 2\eta _{\widehat{a}\widehat{b}}\\
\Gamma _{11} & = & \Gamma _{0}\ldots \Gamma _{9}=\left( \begin{array}{cc}
1_{16} & \\
 & -1_{16}
\end{array}\right) \\
\Gamma _{\widehat{a}}^{T}=-\widehat{C}^{-1}\: \Gamma _{\widehat{a}}\: \widehat{C}\: \: \:  & \widehat{C}^{T}=-\widehat{C}\: \: \:  & \widehat{C}^{\dagger }=\widehat{C}^{-1}
\end{eqnarray*}
 
\item 1+4D (\( AdS_{5} \)) gamma matrices 
\begin{eqnarray*}
\{\gamma _{a},\gamma _{b}\} & = & 2\eta _{ab}\\
\gamma _{0}\gamma _{1}\gamma _{2}\gamma _{3}\gamma _{4} & = & 1_{4}\\
\gamma _{a}^{T}=+C^{-1}\: \gamma _{a}\: C\: \: \:  & C^{T}=-C\: \: \:  & C^{\dagger }=C^{-1}
\end{eqnarray*}
Moreover we use the following (1+4)D \( \gamma  \) explicit representation
\end{itemize}

\[
\gamma _{p}=\left( \begin{array}{cc}
 & \sigma _{p}\\
\widetilde{\sigma _{p}} & 
\end{array}\right) \; \gamma _{4}=\left( \begin{array}{cc}
i\, 1_{2} & \\
 & -i\, 1_{2}
\end{array}\right) \; C=\left( \begin{array}{cc}
i\sigma _{2} & \\
 & i\sigma _{2}
\end{array}\right) \]

with \( p,q,\ldots =0\ldots 3 \), \( \sigma _{p}=\left\{ 1_{2},-\sigma _{1},-\sigma _{2},-\sigma _{3}\right\}  \)
and \( \widetilde{\sigma }_{p}=\left\{ 1_{2},\sigma _{1},\sigma _{2},\sigma _{3}\right\}  \)

\begin{itemize}
\item 0+5D (\( S_{5} \)) gamma matrices
\begin{eqnarray*}
\{\tau _{i},\tau _{j}\} & = & 2\eta _{ij}\\
\tau _{5}\tau _{6}\tau _{7}\tau _{8}\tau _{9} & = & i\: 1_{2}\\
\tau _{i}^{T}=+C^{-1}\: \tau _{i}\: C & \: \: \:  & C_{5}=C\: \: \: 
\end{eqnarray*}

\item 1+9D gamma expressed through \( AdS_{5} \) \( \gamma ^{a} \) and \( S_{5} \)
\( \tau ^{i} \) gammas
\begin{eqnarray*}
\Gamma \widehat{^{a}} & = & \left\{ \gamma ^{a}\otimes 1_{4}\otimes \sigma _{1}\, ,\, 1_{4}\otimes \tau ^{i}\otimes (-\sigma _{2})\right\} \\
\widehat{C} & = & C\otimes C_{5}\otimes \sigma _{2}
\end{eqnarray*}

\end{itemize}
.

\section{The superstring action on \protect\( AdS_{5}\bigotimes S_{5}\protect \) in
first order formalism and its \protect\( \kappa \protect \) symmetry.\label{app-action-first-order}}

It is purpose of this appendix to determine the form of the superstring action
in first order formalism and its \( \kappa  \) symmetry: this is done at the
same time starting from the ansatz inspired by the knowledge of the second order
action (\cite{GHMNT}) and from the fact that the dilatons are constant:
\begin{eqnarray}
S & = & \int _{\Sigma }c_{1}\Pi ^{\widehat{a}}_{\alpha }\: i_{*}V^{\widehat{b}}\wedge e^{\beta }\: \eta _{\widehat{a}\widehat{b}}\: \epsilon _{\alpha \beta }+c_{2}\Pi ^{\widehat{a}}_{\alpha }\Pi ^{\widehat{b}}_{\beta }\: \eta _{\widehat{a}\widehat{b}}\: \eta ^{\alpha \beta }\: \epsilon _{\gamma \delta }\: e^{\gamma }\wedge e^{\delta }\nonumber \\
 &  & +c_{3}V_{+}*A-c_{3}^{*}V_{-}*A\label{S-primo-ordine} 
\end{eqnarray}
where \( V_{\pm }*A=\epsilon _{\alpha \beta }V_{\pm }^{\alpha }A^{\beta } \),
\( \alpha ,\beta ,\ldots \in \left\{ 0,1\right\}  \) are worldsheet indices
and \( i_{*} \) is the pullback on the string worldsheet \( \Sigma  \) due
to the superimmersion \( i:\: \Sigma \: \rightarrow \: AdS_{5}\bigotimes S_{5} \).

The relative coefficient of the first two terms of the first line in eq. (\ref{S-primo-ordine})
is determined to be
\begin{equation}
\label{c2}
c_{2}=-\frac{1}{4}c_{1}
\end{equation}
by the request that the \( \Pi _{\alpha }^{\widehat{a}} \) equation of motion
yields
\begin{equation}
\label{eq-mot-Pi}
\frac{\delta S}{\delta \Pi }=0\Longrightarrow i_{*}V^{\widehat{a}}=\Pi ^{\widehat{a}}_{\alpha }e^{\alpha }
\end{equation}

The zweibein equation of motion gives the Virasoro constraints:
\begin{equation}
\label{Virasoro-const}
\frac{\delta S}{\delta e}=0\Longrightarrow \Pi _{\alpha }\bullet \Pi _{\beta }=\frac{1}{2}\eta _{\alpha \beta }\Pi ^{2}
\end{equation}

From the request of the action (\ref{S-primo-ordine}) (in the 1.5 formalism)
to have a \( \kappa  \) symmetry we determine the relative value of the coefficients
of last two terms in eq. (\ref{S-primo-ordine}) with respect the first term
in the first line: 
\begin{equation}
\label{c3}
c_{3}=\frac{1}{4}e^{i\phi }c_{1}\; \; \; \; \forall \phi 
\end{equation}
and the vector field \( \overrightarrow{\epsilon }=\delta _{\kappa }\Theta _{N}\overrightarrow{\partial _{\Theta _{N}}}+\delta _{\kappa }x\overrightarrow{\partial _{x}} \)
associated to the \( \kappa  \) symmetry and the variation of the zweibein:
\[
\delta _{\kappa }e^{\alpha }=i\overline{\kappa }^{N\alpha }\psi _{N}-i\overline{\psi }^{N}\kappa _{N}^{\alpha }\]
. The vector field \( \overrightarrow{\epsilon } \) is defined by the properties

\begin{eqnarray}
_{\overrightarrow{\epsilon }}\mid \psi _{N} & = & \epsilon _{N}=\Pi ^{a}_{\alpha }\: \gamma _{a}\kappa ^{\alpha }_{N}-i\Pi ^{i}_{\alpha }\: J_{i}|^{P}_{N}\: \kappa ^{\alpha }_{P}\label{kappa-symmetry} \\
\kappa _{c\alpha }^{N} & = & -ie^{i\phi }\: L^{NP}\: \epsilon _{\alpha \beta }\: \kappa _{P\beta }\label{vincolo-su-kappa} 
\end{eqnarray}
and its action by mean of the Lie derivative gives the supersymmetric variations
(up to local Lorentz and gauge transformations) of the other fields 
\begin{eqnarray}
\delta _{\kappa }V^{a} & = & \frac{i}{2}\left( \overline{\epsilon }^{N}\gamma ^{a}\psi _{N}-\overline{\epsilon }_{cN}\gamma ^{a}\psi _{c}^{N}\right) \label{kappa-var-1} \\
\delta _{\kappa }V^{i} & = & \frac{1}{2}\: J^{i}|_{N}^{M}\: \left( -\overline{\epsilon }^{N}\psi _{M}+\overline{\epsilon }_{cM}\psi _{c}^{N}\right) \\
\delta _{\kappa }(V_{+}*A) & = & 2\: \overline{\epsilon }_{cN}\gamma ^{a}\psi _{M}\: L^{NM}\: V_{a}+2i\: \overline{\epsilon }_{cN}\psi _{M}\: J^{i}|^{NM}\: V_{i}\label{kappa-var-5} 
\end{eqnarray}
where we have defined
\begin{eqnarray}
J^{i}|_{N}^{M} & = & \overline{\eta }_{N}\tau ^{i}\eta ^{M}=\overline{\eta }_{c}^{M}\tau ^{i}\eta _{cN}\label{j-inizio} \\
J^{i}|^{MN} & = & \overline{\eta }_{c}^{N}\tau ^{i}\eta ^{M}=-J^{i}|^{NM}\\
J^{i}|_{MN} & = & \overline{\eta }_{M}\tau ^{i}\eta _{cN}=-J^{i}|_{NM}\\
L^{MN} & = & \overline{\eta }_{c}^{M}\eta ^{N}=-L^{NM}\\
L_{MN} & = & \overline{\eta }_{M}\eta _{cN}=-L_{NM}\label{l-fine} 
\end{eqnarray}
It is important in proving that the action (\ref{S-primo-ordine}) is invariant
under the \( \kappa  \) symmetry variations given by eq.s (\ref{kappa-symmetry},\ref{kappa-var-1}-\ref{kappa-var-5})
to use the following Fierz identities:
\begin{eqnarray}
J^{i}|_{N}^{M}\: L^{NP} & = & L^{MP}\: J^{i}|_{P}^{N}=J^{i}|^{MN}\label{fierz-1} \\
L^{MN}\: L_{NP} & = & \delta ^{M}_{P}\\
J^{i}|_{N}^{M}\: J^{j}|_{P}^{N}+J^{j}|_{N}^{M}\: J^{i}|_{P}^{N} & = & 2\eta ^{ij\: }\: \delta ^{M}_{P}\label{fierz-3} 
\end{eqnarray}
The derivation of these Fierz identities is based on the fact that \( \eta ^{N} \)
are a basis of the vector space on which the spinor irrep acts, i.e. \( \overline{\eta }_{N}\tau \: \eta ^{N}\propto tr(\tau ) \)
.

Before closing this appendix we would like to make some comments:

\begin{itemize}
\item here we have 4 complex \( \kappa _{N\alpha } \) parameters instead of the usual
real two because our superspace is five dimensional and the \( \psi _{N} \)
(\( \Theta _{N} \)) are the dimensional reduction of the usual 10D Majorana
\( \Psi  \) (\( \Theta  \)) hence they transform as spinors under \( SO(1,4) \)
\emph{only}; 
\item Eq. (\ref{vincolo-su-kappa}) is the analogous of the (anti)selfduality condition
imposed on the \( \kappa _{\alpha } \) parameter in the usual GS superstring;
\item even if it is not immediate to see, the \( \psi _{N} \) contraction involves
a projector applied on the \( \kappa  \) parameters. We can in fact define
\begin{equation}
\label{projector+-}
{\cal P}^{(\pm )}_{\alpha }|_{N}^{M}=\Pi ^{a}_{\alpha }\: \gamma _{a}\: \delta ^{M}_{N}\pm i\: \Pi ^{i}_{\alpha }\: J^{i}|_{N}^{M}
\end{equation}
and verify the following properties
\begin{eqnarray*}
{\cal P}^{(+)}_{\alpha }|_{N}^{M}+{\cal P}^{(-)}_{\alpha }|_{N}^{M} & = & 2\Pi ^{a}_{\alpha }\: \gamma _{a}\: \delta ^{M}_{N}\\
{\cal P}^{(-)}_{\pm }|_{N}^{M}{\cal P}^{(+)}_{\pm }|_{P}^{N} & = & 0\\
{\cal P}^{(s)}_{\pm }|_{N}^{M}{\cal P}^{(s)}_{\pm }|_{P}^{N} & = & 2\Pi ^{a}_{\pm }\: \gamma _{a}\: {\cal P}^{(s)}_{\pm }|_{P}^{M}
\end{eqnarray*}
which shows as the \( {\cal P} \)s be a projectors on the space where \( \Pi ^{a}_{\pm }\: \gamma _{a} \)
is invertible, i.e. when \( \Pi ^{a}_{\pm }\: \eta _{ab}\Pi ^{b}_{\pm }\neq 0 \).
Moreover we have 
\begin{equation}
\label{explicit-projector-on-k}
\epsilon _{N}\pm ie^{-i\phi }L_{NP}\epsilon ^{P}_{c}=2{\cal P}^{(-)}_{\pm }|_{N}^{M}\kappa _{M\mp }
\end{equation}
which clearly shows that at least half the degrees of freedom associated with
\( \kappa _{N} \) are projected out.
\end{itemize}

\section{How the supersolvable algebra fixes the \protect\( \kappa \protect \) symmetry.\label{Why-kappa-fix}}

In order to discuss and explain the issue of why the \( \kappa  \) symmetry
is properly fixed which by the gauge which is automatically chosen by the supersolvable
algebra we can either work using the general formulae derived in second order
formalism in (\cite{GHMNT}) or use the formulae written in first order formalism
in the previous app. \ref{app-action-first-order}: in both cases where we can
easily find the gravitino variations but there is a big difference between the
two results since while the formulae in ref. \cite{GHMNT} refer to the 10D
gravitini \( \Psi  \) the ones in the previous appendix refer to the dimensional
reduced ones \( \psi _{N} \) with which we have been working in this paper.
We will hence follow this second approach.

As a first step we write down the supersolvable projector associated with the
fermionic coordinates which are projected out:

\[
P=\frac{1+i\gamma _{4}}{2}\]
 and we notice that the same projector \( P \) has to be used to find the gravitini
components which are projected out. In the following we find under which conditions
we can find a vector field \( \overrightarrow{\epsilon } \) which can be used
to eliminate the wanted components of the gravitini \( \psi _{N} \) .

To this purpose we write down the contraction of the vector field \( \overrightarrow{\epsilon } \)
generating the \( \kappa  \) symmetry on \( P\psi _{N}=\psi _{N}^{(+)} \)
and \( (P\psi _{N})_{c}=Q\psi _{c}^{N}=\psi _{c}^{N(-)} \) where \( Q=\frac{1-i\gamma _{4}}{2} \).
After the elimination of \( \kappa _{N1} \) using eq. (\ref{vincolo-su-kappa})
eq. (\ref{kappa-symmetry}) becomes :

\begin{eqnarray}
\epsilon _{N}^{(+)} & = & {\cal A}_{0}|_{N}^{M}\: \lambda _{M}+ie^{-i\phi }{\cal C}_{1}|_{N}^{M}\: L_{MP\: }\: \lambda _{c}^{P}+\nonumber \\
 &  & +{\cal C}_{0}|_{N}^{M}\: \mu _{M}+ie^{-i\phi }{\cal A}_{1}|_{N}^{M}\: L_{MP}\: \mu _{c}^{P}\label{espilon} \\
ie^{-i\phi }L_{NP}\: \epsilon _{c}^{P(-)} & = & {\cal A}_{1}|_{N}^{M}\: \lambda _{M}+ie^{-i\phi }{\cal C}_{0}|_{N}^{M}\: L_{MP}\: \lambda _{c}^{P}\nonumber \\
 &  & +{\cal C}_{1}|_{N}^{M}\: \mu _{M}+ie^{-i\phi }{\cal A}_{0}|_{N}^{M}\: L_{MP}\: \mu _{c}^{P}\label{L-epsilon-c} 
\end{eqnarray}
where we have defined \( \mu _{M}=P\kappa _{M0} \) , \( \lambda _{M}=Q\kappa _{M0} \)
and
\begin{eqnarray*}
{\cal A}_{\alpha }|_{N}^{M} & = & \Pi ^{p}_{\alpha }\: \gamma _{p}\: \delta ^{M}_{N}\\
{\cal C}_{\alpha }|_{N}^{M} & = & \Pi ^{4}_{\alpha }\: \gamma _{4}\: \delta ^{M}_{N}-i\: \Pi ^{i}_{\alpha }\: J^{i}|_{N}^{M}
\end{eqnarray*}
We can now make the following linearly independent combinations:

\begin{eqnarray}
\Sigma _{N}=\epsilon _{N}^{(+)}+ie^{-i\phi }L_{NP}\epsilon _{c}^{P(-)} & = & \left( {\cal A}_{0}+{\cal C}_{1}\right) |_{N}^{M}\left[ \lambda _{M}+ie^{-i\phi }L_{MP}\lambda _{c}^{P}\right] +\nonumber \\
 &  & +\left( {\cal A}_{1}+{\cal C}_{0}\right) |_{N}^{M}\left[ \mu _{M}+ie^{-i\phi }L_{MP}\mu _{c}^{P}\right] \label{eq-sigma} \\
\Delta _{N}=\epsilon _{N}^{(+)}-ie^{-i\phi }L_{NP}\epsilon _{c}^{P(-)} & = & \left( {\cal A}_{0}-{\cal C}_{1}\right) |_{N}^{M}\left[ \lambda _{M}-ie^{-i\phi }L_{MP}\lambda _{c}^{P}\right] +\nonumber \\
 &  & -\left( {\cal A}_{1}-{\cal C}_{0}\right) |_{N}^{M}\left[ \mu _{M}-ie^{-i\phi }L_{MP}\mu _{c}^{P}\right] \label{eq-delta} 
\end{eqnarray}
 Since these combinations are linearly independent also on the complex field
because under charge conjugation each of them transforms in its selves we can
consider them independently and we can, for example, try to solve the \( \Sigma  \)
equation for either \( \left[ \lambda _{M}+ie^{-i\phi }L_{MP}\lambda _{c}^{P}\right]  \)
or \( \left[ \mu _{M}+ie^{-i\phi }L_{MP}\mu _{c}^{P}\right]  \).

We find that in both cases the necessary condition for the invertibility of
the matrices \( \left( {\cal A}_{0}+{\cal C}_{1}\right)  \) and \( \left( {\cal A}_{1}+{\cal C}_{0}\right)  \)
is (using the Virasoro constraint eq. (\ref{Virasoro-const}) that
\begin{equation}
\label{kappa-descrive-app}
\Pi ^{p}_{\tau }\Pi _{p\tau }+\Pi ^{4}_{\sigma }\Pi _{4\sigma }+\Pi ^{i}_{\sigma }\Pi _{i\sigma }\neq 0
\end{equation}
One could wonder whether this is a sufficient condition for the solvibility
of eq.s (\ref{eq-sigma},\ref{eq-delta}), the answer is yes since would the
sum of the images of the two matrices \( \left( {\cal A}_{0}+{\cal C}_{1}\right)  \)
and \( \left( {\cal A}_{1}+{\cal C}_{0}\right)  \) be the whole space spanned
by \( \Sigma  \) then it should be possible to find a number \( \alpha  \)
such that \( \left( {\cal A}_{0}+{\cal C}_{1}\right) +\alpha \left( {\cal A}_{1}+{\cal C}_{0}\right)  \)
is always invertible but it is easy to verify that the determinant of the previous
combination is proportional to eq. (\ref{kappa-descrive-app}) too.

We conclude therefore that states which do not satisfy eq. (\ref{kappa-descrive-app})
cannot be described in this gauge.


\begin{thebibliography}{}\bibitem{Maldacena}J. Maldacena, The Large \( N \) Limit of Superconformal Filed Theory and Supergravity,
\hepth{9711200}
\bibitem{GubserKlenanovPolyako}S.S. Gubser, I.R. Klebanov and A.M. Polyakov, Gauge Theory Correlators from
Non-Critical String Theory, \hepth{9802109}
\bibitem{Witten}E. Witten,Anti de Sitter Space and Holography, \hepth{9802150}
\bibitem{MatsaevTseytlin}R.R Metsaev and A.A. Tseytlin, Type IIB Superstring Action in \( AdS_{5}\times S_{5} \)
Background, \hepth{9805028}; \\
Supersymmetric D3 Brane Action \( AdS_{5}\times S_{5} \), \hepth{9506095}
\bibitem{Kallosh}R. Kallosh, J. Rahmfeld and A. Rajaraman, Near
  Horizon Superspace, \hepth{9805217}
\bibitem{Kall-fixing}R. Kallosh, Superconformal Actions in Killing
  Gauge, \hepth{9807206}
\bibitem{IIB}L. Castellani and I. Pesando, The Complete Superspace
  Action of Chiral \( D=10 \) \( N=2 \) Supergravity, \ijmpa{8}{1993}{1125};\\
L. Castellani, Chiral \( D=10 \),\( N=2 \) Supergravity on the Group Manifold:
1. Free Differential Algebra and Solution of Bianchi Identities ,
\npb{294}{1987}{877}.
\bibitem{gruppoTo}G. Dall'Agata, D. Fabbri, C. Fraser, P. Fre', P. Termonia and M. Trigiante,
The \( Osp(8|4) \) Singleton Action from the Supermembrane, \hepth{9807115}
\bibitem{GHMNT}M.T. Grisaru, P. Howe, L. Mezincescu, B.E.W. Nilsson and P.K. Townsend, \( N=2 \)
Superstrings in a Supergravity Background, \plb{162}{1985}{116}.
\bibitem{D3-to}F. Cordaro, L. Gualtieri and I. Pesando, The \( \kappa  \) Gauge Fixed Action
for the \( D3 \) brane and its singleton, to appear.
\bibitem{BanksGreen}T. Banks and M.B. Green, Non-perturbative Effects in \( AdS_{5}\bigotimes S_{5} \)
String Theory and d=4 SUSY Yang-Mills, JHEP05 (1988) 002, \hepth{9804170}
\bibitem{IIB-HW}P.S. Howe and P.C. West, The Complete \( N=2 \) \(
  d=10 \) Supergravity, 
\npb{238}{1984}{ 181}
\bibitem{Li}M. Li, Evidence for Large N Phase Transition in \( N=4 \) Super Yang-Mills
Theory at Finite Temperature, \hepth{9807196}
\end{thebibliography}
\end{document}